# The Quantified Community at Red Hook: Urban Sensing and Citizen Science in Low-Income Neighborhoods

Constantine E. Kontokosta, PhD, PE
NYU Center for Urban Science and Progress & Tandon School of Engineering
Brooklyn, NY, USA
ckontokosta@nyu.edu

Nicholas Johnson
University of Warwick & NYU Center for Urban Science and Progress
Brooklyn, NY, USA
Nicholas.johnson@nyu.edu

Anthony Schloss
Red Hook Initiative
Brooklyn, NY, USA
tony@rhicenter.org

## ABSTRACT

The Quantified Community (QC)—a long-term neighborhood informatics research initiative—is a network of instrumented urban neighborhoods that collect, measure, and analyze data on physical and environmental conditions and human behavior to better understand how neighborhoods and the built environment affect individual and social well-being. This initiative is intended to create a data-enabled research environment to rigorously study the complex interactions in urban neighborhoods. The QC has initially launched in three very distinct areas in New York City: at Hudson Yards, a ground-up "city-within-a-city" of approximately 20 million square feet in Manhattan, in collaboration with the Related Companies; in Lower Manhattan, a mixed-use neighborhood that attracts residents, workers, and visitors, in collaboration with the Alliance for Downtown NY; and in Red Hook, Brooklyn, an economically-distressed community facing significant development and demographic changes, in partnership with the Red Hook Initiative. This paper describes our recent pilot project to deploy novel urban sensors in Red Hook to collect and analyze quality-of-life measurements at high spatial and temporal resolution. This effort is complemented by a citizen science initiative to engage local residents in the data collection and problem-solving process to drive evidenced-based community decision-making and improve local and city governance.

## 1. INTRODUCTION

The Quantified Community (QC)—a long-term neighborhood informatics research initiative—is a network of instrumented urban neighborhoods that collect, measure, and analyze data on physical and environmental conditions and human behavior to better understand how neighborhoods and the built environment affect individual and social well-being (Kontokosta 2016). To advance this work, a new urban sensing platform has been developed, the QC Urban QoL Sensor—a low-cost, but reliable sensor array to measure and track localized environmental conditions, down to the individual street, block, or building. The devices measure air quality, noise, light levels, pedestrian counts, and temperature/pressure/humidity. These sensor data are combined with administrative, mobility, social media, and Wi-Fi usage data to create a neighborhood profile to benchmark changes over time and compare to other areas of the city. These data are designed to help communities identify and solve problems, focusing on issues of environmental health and mobility, through new sensing modalities, analytics, and data visualization that enable a deeper partnered with the Red Hook Initiative (RHI), a local social services community organization, to install the sensors and engage with the local community to provide additional, volunteered data. As part of the QC initiative, we also work closely with the NYC Mayor's Office of Technology and Innovation and the NYC Economic Development Corporation on aspects of this work in other parts of the City. Data transparency and accessibility are paramount to this research, and we are developing tools and citizen science programs to disseminate the data and analysis and to directly engage local community members in the data collection and problem-solving components of the research.

The neighborhood presents a promising locus to more fully understand complex urban systems and the diversity of interactions within and across heterogeneous communities (Kontokosta 2016). Building on the prior work of urbanists, such as William Holly Whyte, whose synoptic observation of urban life provided a means to understand social interactions and the mediating role of urban space (Whyte 1980), the QC research project leverages advancements in sensing technologies and computational efficiencies to expand the scope and nature of what is knowable about the patterns of urban life.

Our motivation stems from how little is known about how the built environment and urban environmental conditions impact individual and community well-being and health. Ultimately, this research can support a more complete understanding of the interaction of poverty, health, and the environment in urban neighborhoods. The primary objective is to increase the spatial and temporal resolution of data collection to better identify, and respond to, the triggers of urban environmental stressors and diminished quality-of-life. The intensive study of neighborhoods will create the observational and participatory data to build data-driven models of local urban environments and the causes and effects of various outcomes. In addition, this informatics infrastructure creates a long-term study environment to measure and benchmark how the social, economic, physical, and environmental characteristics of neighborhoods change over time. The use of data technologies in this way can help us examine how city residents are impacted by the neighborhood and buildings in which they live, and to measure and evaluate the impact of changes in built, natural, and social systems over time.

Through both short- and long-term milestones, the QC project is expected to achieve the following outcomes:

- Create new models and metrics for sustainable urban development and placemaking

- Develop rigorous, objective metrics to evaluate cities and development across sustainability, resilience, and health dimensions to support a global transfer of lessons learned

- Test, refine, and develop new civic technologies Develop individual outcomes across measures of health, activity, productivity, resilience, and resource conservation

- Redefine the science and practice of community planning and urban design through evidence-based research

CUSP has initially launched the QC in three very distinct neighborhoods in New York City: at Hudson Yards, a ground-up "city-within-a-city" of approximately 20 million square feet in Manhattan, in collaboration with the Related Companies; in Lower Manhattan, a mixed-use neighborhood that attracts residents, workers, and visitors, in collaboration with the Alliance for Downtown NY; and in Red Hook, Brooklyn, an economically-distressed community facing significant development and demographic changes, in partnership with the Red Hook Initiative. In each of these communities, we are working with different constituents to define contextual problems and collect relevant data. This paper describes our recent pilot project in Red Hook to deploy original urban sensors to collect and analyze quality-of-life measurements at high spatial and temporal resolution. This effort is complemented by a citizen science initiative to engage local residents the in data collection and problem-solving process to drive evidenced-based community decision-making.

## 2. APPLICATIONS AND USE CASES OF HIGH-RESOLUTION URBAN DATA IN THE COMMUNITY CONTEXT

Cities can be viewed as the dynamic interactions of numerous complex sub-systems, each with their own rhythm, function, and logic. As Meadows (2002) illustrated, understanding system behavior requires a combination of patient observation and learning, an appetite to question and challenge models, and an appreciation of 'messiness'. Cities are in many ways the 'messiest' of complex systems, blending human, social, environmental, and physical systems within the constraints of regulatory, economic, and political realities (Bettencourt 2014). The challenge of understanding this complex system is all the more urgent, given the rapid pace of global urbanization and the social challenges that have accompanied this growth. It is not sufficient to simply model the characteristics and behavior of the urban environment; scientists must collect and analyze massive, novel, and diverse data of appropriate spatial and temporal resolution so individuals, communities, and policy-makers can decide how best to intervene to improve quality-of-life.

A case in point: the neighborhood of Red Hook experiences an asthma rate of more than 2.5 times that of the national average (NYC DOHMH 2015). Its residents, with more than a third living below the federal poverty line, can expect to live approximately ten fewer years than those in New York City's wealthiest communities. By collecting air quality data - particularly PM2.5 concentrations, which have been shown to have a direct impact on cardiovascular and cardiopulmonary health (Pope III and Dockery 2006) - at the block level, it is possible to assess both the causes and the impact of poor air quality on resident health outcomes with unprecedented detail (Gao, Cao, and Seto 2015). Coupling these data with building energy use data, for example, begins to allow us to empirically explore how the quality of the built environment, human behavior, and environmental conditions interact to affect public health.

Another use case relates to what is known as the urban heat island (UHI), an intensely studied phenomena in which man-made urban surfaces (concrete, asphalt, etc.) lead to an increased heating of the surrounding area (Oke 1973; Oke 1982; Kim 1992; Arnfield 2003). The urban heat island effect has also been linked with serious health impacts and is expected to continue to affect the growing megacities around the world (Tan et al. 2010; Mavrogianni et al. 2011). By measuring temperature at high spatial and temporal resolution (real-time, at the block scale), the implications of the urban heat island on issues such as individual health outcomes and building energy consumption can be examined. These measurements can also identify buildings and communities with high vulnerability to heat waves and other heat-related emergencies.

Ultimately, there is a need to explore both the correlations and causal relationships between various aspects of the phsyical, environmental, and social components of the community. For example, how do light levels impact crime activity? Do persistent noise levels affect stress levels and health outcomes for those exposed? Are there spatial and temporal patterns in air quality measurements that can lead to understanding the cause of high levels of particulates? Does land use impact mobility, and how is this in turn affected by weather? Answers to these questions, which could provide new insights into the urban environment, require a level of measurement and analysis not currently practiced in cities.

In addition to neighborhood planning, real-time data collection, integration, and analysis can provide efficiencies in public service delivery. For instance, the real-time monitoring of intersection and street-level temperature measurements can provide an early warning indicator for vulnerable populations in the case of extreme heat. Through data feedback mechanisms, areas of high particulate concentrations can be identified, and those with respiratory illnesses can be alerted. The use of real-time analytics, coupled with causal analyses that are not limited to predictive models, can provide new insights to support both operational and longer-term policy and planning goals.

### 2.1. CITIZEN SCIENCE, PARTICIPATORY SENSING, AND COMMUNITY ENGAGEMENT

The importance of community engagement to the future of urban science is non-trivial; to link scientific modeling and analysis approaches to real-world impact requires both an understanding of the problems faced by communities and the cooperative analysis of data. This then connects fundamental knowledge gained with the "ground truth" of the people who have a unique understanding of the actual conditions that can be lost in model abstractions.

We see several opportunities to directly engage communities and local residents in urban informatics and data science through the QC project. First, we use citizen science initiatives to bring the community into the process of problem identification and problem-solving through participatory data collection and analysis (Elwood 2002). By involving residents in the research, and allowing them to help shape the hypotheses to be tested, the scope of possible research questions increases (Bonney et al. 2009).

Second, participatory sensing creates opportunities for residents to provide data that they feel may further the understanding of their own neighborhood. Through the QC initiative, we have initiated a project we call CommunitySense that uses Physical Web devices to allow smart-phone users to interact with our sensors. When properly configured, if a person walks past one of our sensors with a Bluetooth-enabled device, a URL is presented on the device screen. This URL takes the user to a website that shows data collected by that sensor, and how it compares to other sensor data from the neighborhood. In addition, we are developing a web-based, interactive survey that will measure the difference between subjective and objective measurements of quality-of-life indicators (e.g. heat) to understand how the perception of different environmental conditions varies across individuals and neighborhoods.

Third, community engagement beyond data collection and problem-solving is necessary to drive change. Urban data and technology must be viewed as an enabler of improved decision-making by residents, communities, and city policymakers more broadly. Understanding local problems, and how they can be solved, requires a deep appreciation of a community, its people, its history, and its future goals.

Ultimately, transparency in the collection, use, and dissemination of data and subsequent analysis is necessary to build trust with local communities and to minimize the potential for unintended abuses of sensitive data. The data protocols for the QC require that all data collected, and its intended use, are made known to local stakeholders and made available at an appropriate level of aggregation to protect sensitive data, while still communicating a representative sample of information. By establishing a dynamic flow of information with the community, the various stakeholders in the QC can more effectively work together to address identified challenges.

## 3. URBAN SENSING IN LOW-INCOME COMMUNITIES

Quantification of urban systems is a critical tool to understand cities and can provide enhanced decision-making abilities useful for planning, benchmarking and evaluating urban systems. A necessary component of quantification are the sensors and sensor networks that have received increased attention in recent years (Yick, Mukherjee, and Ghosal 2008; Chong and Kumar 2003). The proliferation of low-cost digital technologies and increased connectivity creates a new capacity to collect real-time, in-situ information to better understand the complex interactions between urban ecosystems and infrastructure.

Urban Internet of Things (IoT) technologies are viewed as the next generation of urban infrastructure. However, low-income and economically distressed communities are often not the focus of urban and civic technology deployments, and the question of equity continues to be overshadowed by financial expedience in technology-led city strategies. These communities tend not to have the necessary physical infrastructure to support such technologies, and the limited access to, and awareness of, information and communication technologies have hampered innovation in neighborhoods that could benefit substantially from greater access to information and improved public service delivery.

While sensor networks are often used in cities, individual network nodes often focus on quantifying only one aspect of the environment and place little emphasis on developing a low-cost platform. The impetus for low-cost sensing stems from the potential for distributed sensor deployments to provide information with increased spatial and temporal granularity. In addition, low-cost sensing platforms provide an opportunity to engage communities typically excluded from many of the "Smart City" discussions, those that lack the financial resources to support technology installations or the access to connectivity needed for many IoT deployments. Given this, and the fact that sensing technologies are evolving rapidly, the focus of our work is on non-invasive, cost-effective, but reliable, urban sensor nodes.

The design of a low-cost, reproducible and interoperable sensor platform presents a range of technical challenges. Real-world deployment, however, presents an entirely new, additional set of difficulties, including site selection, mitigating impacts of external elements (i.e. weather or human tampering with the device), and legal and privacy issues inherent in collecting information in urban spaces. These elements combine to make sensor deployment in urban spaces challenging and require close community relationships in order to facilitate successful sensor implementations.

A potential trade-off when using low-cost solutions are limitations in data accuracy and resolution. However, the desired accuracy and resolution depends on the specific question or use case being addressed and in many cases, relative readings provide useful information when absolute readings are not available.

We are purposely developing and testing low-cost, non-invasive sensor arrays for use in the QC instrumentation environment. With the rapid evolution of sensor technology, inexpensive sensors that can be easily installed, removed, and adapted to changing needs and conditions provide an optimal degree of flexibility and scalability. However, calibration remains a critical element of the early stage proof-of-concept; finding the appropriate balance of cost, adaptability, reliability, and durability in the urban environment is a significant component of the test phase research.

### 3.1. THE QC URBAN QoL SENSOR

Several hardware platforms were considered and tested, though final platform selection was determined by power limitations created by the specific deployment locations. Test locations were initially assumed to be without power and a battery would therefore be necessary. The 5v Trinket Pro manufactured by Adafruit Industries was selected as the final platform for its low power consumption (30mA during normal operation), low cost ($9.95USD) and small size (38.1mm x 17.7mm x 5mm).

The Trinket Pro uses the ATmega328 micro-controller chip which has a 16Mhz clock, 28k FLASH memory, 18 GPIO pins and 8 analog inputs and operates on 5VDC. The Trinket Pro has a 10 bit analog-to-digital converter(ADC) and a voltage regulator capable of regulating up to 16VDC. Data are stored locally using a micro-sd card and real-time clock for time-stamped data.

Baseline community measurements focused on environmental measurements including air quality, temperature, pressure, humidity, and luminosity. Sensor selection was largely based on cost, although significant consideration was given to accuracy, resolution, integration complexity, and power consumption.

In order to provide air quality readings, our sensor collects particle concentrations measurements and can detect particles as small as 0.1μm. Noise levels were measured using an analog microphone has a frequency response of 100Hz - 15kHz, a -42dBv sensitivity and a 62 dBA signal-to-noise ratio. A 3.3v regulator (L4931) was used to provide the microphones required voltage of 3.3VDC. To increase the dynamic range of the ADC, the regulated 3.3v was provided to the Trinket's analog reference (AREF) input. A smoothing filter was applied to the incoming signal because of the overall sampling interval of the sensor platform.

The temperature, pressure and humidity sensor has an accuracy of +/- 0.5C, +/-3% relative humidity and +/- 1hPa. Luminosity was captured by across two channels; the first channel is from 300-1100nm and a second channel from 500-1100nm.

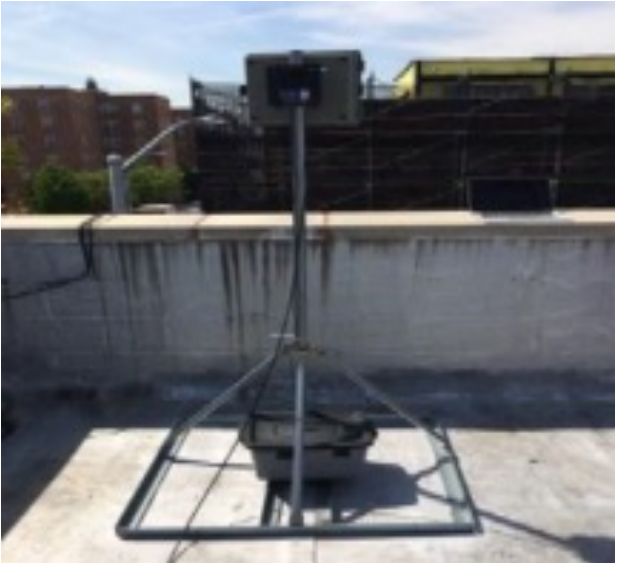

*Fig.1: QC Urban QoL Sensor installed at RHI roof (sensor unit is black; the grey box behind is a Wi-Fi node).*

Overall design of the housing included considerations for weatherproofing, access for data retrieval and durability (see Figure 1). The electronics were enclosed in a modified electrical box made of PVC(114.3mm x 116.8mm x 50.8mm). 20mm holes were cut in the top and bottom. The top hole was sealed with a clear piece of acrylic in order to expose the luminosity sensor. A 10% decrease in overall luminosity readings was observed from the addition of the acrylic. The two holes in the bottom exposed the temperature/pressure/humidity sensor, as well as the microphone. Holes (1.5mm) were cut on in the top plate to expose the intake and outtake of the particulate sensor to the air. Above these holes, two laser-cut pieces of acrylic were bent and glued to the top plate to form an awning preventing water from entering the openings. Figure 2 shows a close-up of the housing and internal components of beta version of the sensor.

Integration of the microphone and the dust sensor into the platform required specific adjustments because of the demand for a high sample rate to provide meaningful data. While a high sample rate can be achieved by the Trinket Pro, the integration of both sensors posed potential for data loss given that the micro-controller can only process one at a time. In order to properly integrate both sensors, a second Trinket Pro was used. The second Trinket Pro was dedicated to noise sampling and was overclocked to provide a sampling rate of 60kHz. The first Trinket Pro used an interrupt to accurately capture LPO readings from the dust sensor.

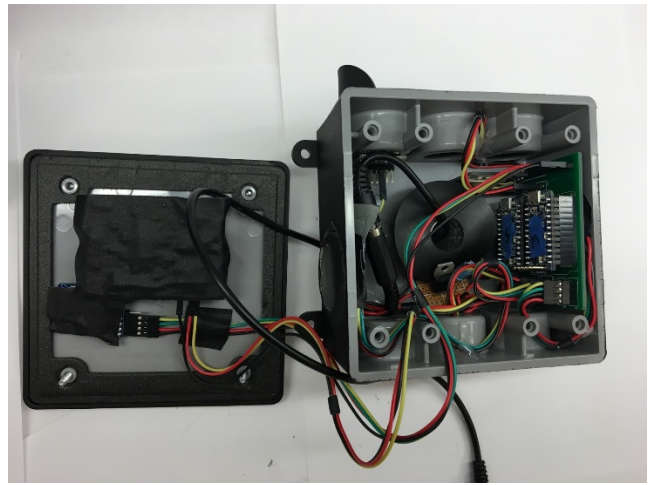

*Fig.2: Beta version of the QC Urban QoL Sensor housing and internal configuration.*

Data were sampled from all sensors every five seconds. During normal operation, overall device power consumption is approximately 130mA. The bulk of this consumption is from the dust sensor which uses 90mA for normal operation while the other sensors' power consumption is nominal.

## 3.2 Initial Results from the QC@Red Hook Pilot Project

Four sensors were installed Red Hook: two at the RHI office (one at ground level and one one the rooftop), one at the RHI TechLab and one at the Pioneer Works building. All four sensors have been running continuously since deployment on June 9, 2016.

Our initial analysis of particulate concentrations (air quality) reveal potentially interesting observations. Figure 3 shows average hourly particulate concentrations for each of the four sensors over a one-week period between June 19th and June 26th. While a regular daily pattern emerges, we observe several anomalous readings, including those that occur across the four locations and those that are isolated to individual sensors. Differences between sensors allow for comparison between different parts of the neighborhood and the establishment of a baseline reading. For example, Figure 3 shows increased dust readings for both RHI Office sensors compared to the other two sensors from June 23rd through June 26th. While source identification of these differences is challenging, the differences highlight spatial variation in dust readings throughout Red Hook. The significant variations in particulate matter support the need for high resolution data collection for air quality assessment, beyond the rather coarse measurements currently used to assess city-wide air quality.

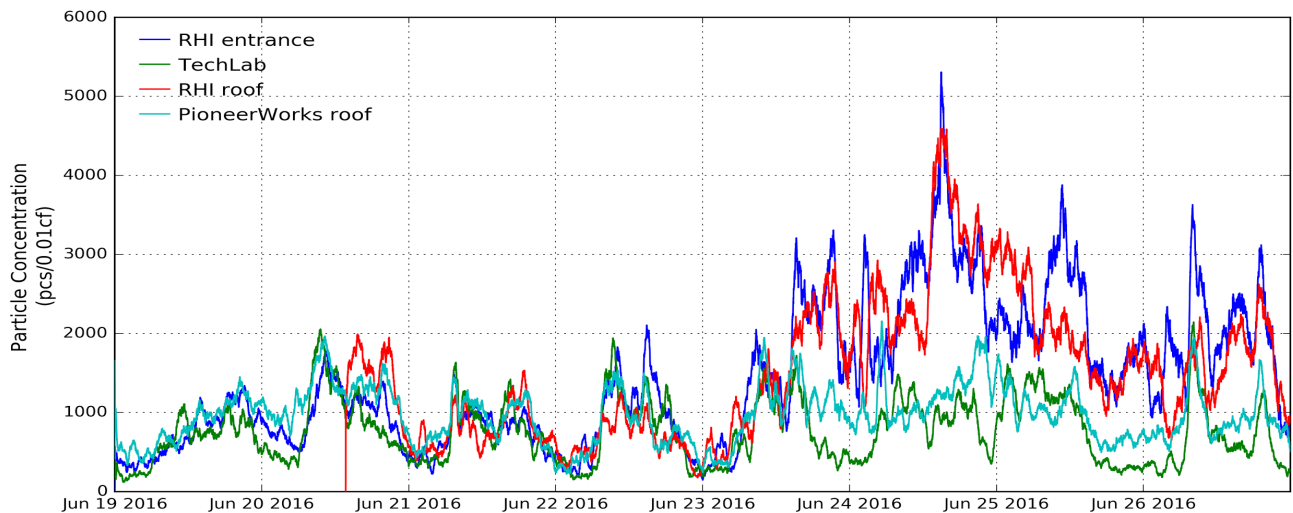

*Figure 3: Average hourly particulate concentrations for four sensors in Red Hook over an eight-day period.*

Potential confounding factors leading to high dust readings are also considered. Temperature and humidity are known factors that can potentially skew readings. Figure 4 shows a scatterplot and regression analysis best-fit line to demonstrate the relationship between humidity and dust. These results indicate a statistically significant positive correlation.

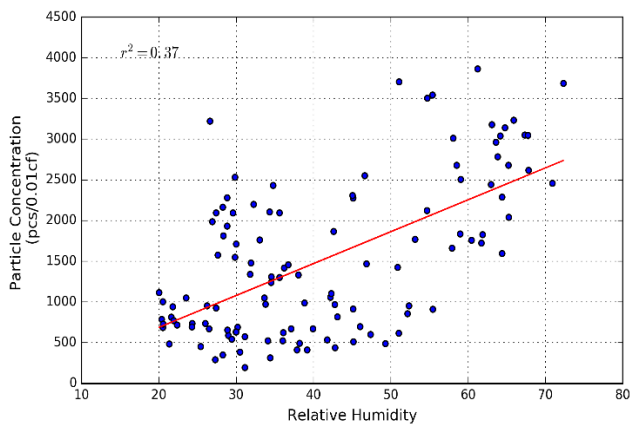

*Figure 4: Scatter plot and regression analysis with relative humidity as the independent variable and dust concentrations as the dependent variable.*

Temperature differentials are noted based on the location of the particular sensor installation. Figure 5 shows the distribution of hourly temperature differences between the sensor installed at the entrance of the RHI office and the one installed on its roof (see Figure 1). Here, we find the roof measurement to be as much as 18 degrees Celsius higher than the street-level measurement during the daytime, while differences in evening readings are less than 1 degree. This difference is impacted by many factors including relative humidity, light levels, and wind speed and direction for a given point in time, and highlights the potential usefulness in localized temperature data and the effects of the built infrastructure on environmental and neighborhood conditions.

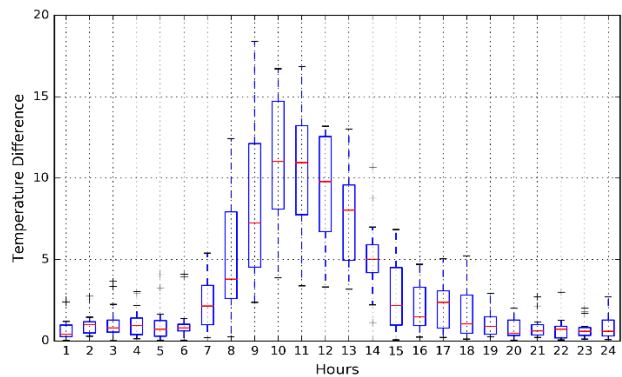

*Figure 5: Distribution of hourly temperature differentials between two sensors installed at the Red Hook offices. One sensor is installed on the roof while the other is installed at ground level.*

Finally, to attempt to (manually) identify a causal relationship between air quality measurements and specific events or activities, we noted unusually high particulate concentrations on Friday, June 10th between 3pm and 6pm shown in Figure 6. The high concentrations were found at both sensors installed at RHI's office, although we find lower values at the rooftop sensor. We asked RHI if anything unusual had occurred that day; it so happens that RHI held a BBQ that afternoon, and our sensors were capturing the particulate matter emanating from the grill. While anecdotal, this example illustrates the potential use of real-time, granular air quality measurements to detect anomalous levels of particulates that might have a direct health impact on those with respiratory illnesses or vulnerable immune systems.

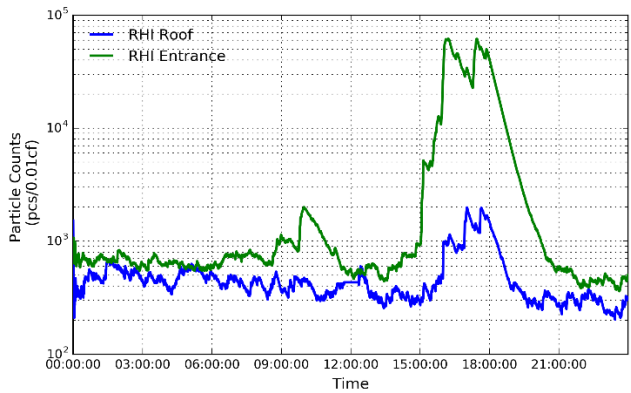

*Figure 6: Particulate readings from RHI offices on June 10th. Both readings identified an anomalous event attributed to a barbeque that took place that day. (note the y-axis logarithmic scale)*

This example also highlights the, perhaps obvious, limitations of manually classifying anomalous events. We are developing unsupervised and semi-supervised machine learning algorithms to classify and label patterns identified in the sensor data. While manual training may be needed, we are building methods to define a "neighborhood signature" or "pulse" that can categorize "normal" activity in a neighborhood, and deviations from it. Such an understanding would help in neighborhood planning, local social and city service delivery, and in emergency response.

Along with our sensor data, we are also collecting and integrating data from heterogeneous sources about the community. This includes city administrative records (land use, building permits, etc.), citizen-provided neighborhood information (311 complaints, etc.), topographic and ecological data (location and canopy size of street trees, ground elevation, etc.) and social media (Twitter, Yelp, etc.). Once merged and geocoded, these data provide a QC baseline for the neighborhood. Together with sensor data, these integrated data provide a robust set of characteristics to model and analyze interactions and relationships across multiple dimensions.

## 3.3 Initial Results from the QC Citizen Science Day

On June 23rd, the team organized a Citizen Science (CS) day together with RHI's Digital Stewards. The RHI Digital Steward program trains young adults from Red Hook to use technology as a pathway to community development and employment. Approximately 20 Digital Stewards participated.

Data were collected using three sensor devices, plus two high-quality, compliance-grade sensors to provide calibration data. These sensors measured temperature (C), noise (dB), and air quality (particulates >1µm per 0.01 cf) at five second intervals. Data were stored locally to a sd card, and the devices were powered by a standard 9V battery. The Digital Stewards were divided into teams, each supported by CUSP students and researchers, and asked to delineate a walking path through the neighborhood that they thought, from their local knowledge, may produce interesting insights from quality-of-life indicator measurements. The team trajectories, tracked using GPS coordinates, are shown in Figure 7.

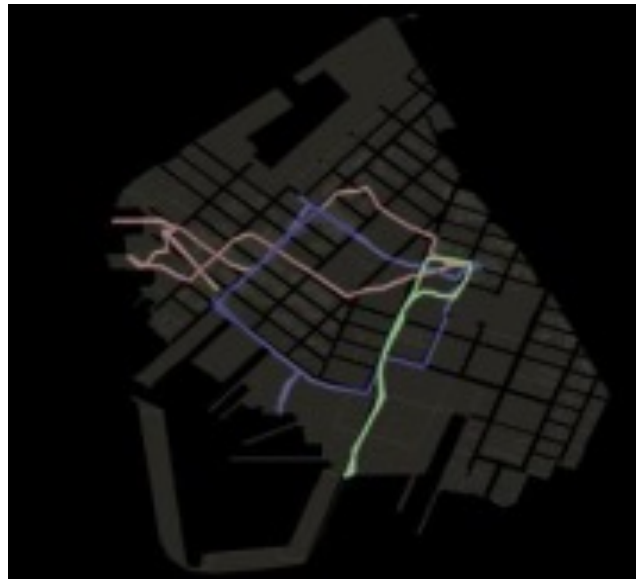

*Fig.7: Citizen science day team walking/data collection trajectories from smartphone GPS coordinates.*

The purpose of the CS day was to (1) test and calibrate low-cost sensors and the potential of participatory sensing initiatives to supplement fixed, in-situ sensor nodes, (2) identify potential "hot-spots" or areas of particularly anomalous measurements to prioritize deployment of future fixed QC sensors, and, most importantly, (3) to engage the local community and residents in the problem identification, data collection, and data analysis process. This objective, community engagement, is critical to impact-driven urban sensing deployments, as working with local residents builds trust, supports data literacy and transparency, and identifies problems that may be addressed through data science by those that know the area best. An example of the collected temperature data by each group is shown in Figure 8.

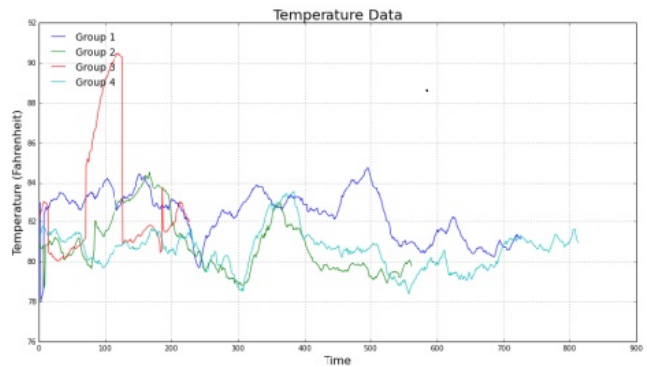

*Fig.8: Citizen science day collected temperature (raw) data, by team and time from start..*

As the teams walked through the neighborhood collecting data, they were asked to note events or activities that may cause unusual readings. These included trucks passing by, construction activity, loud music, and, in one case, a lawnmower. By recording these events, the goal was to see if a correlation could be found in the data being collected. In the discussion that followed the data collection portion of the day, the Digital Stewards were asked to describe other quality-of-life indicators that would be useful to measure, and where there might be particular locations of concern (near the Red Hook concrete plant, for instance) that would be appropriate for more persistent data collection efforts.

A number of lessons were learned during the urban sensing CS day. First, more training is needed to improve data accuracy and reliability. Proper sensor handling techniques, together with a more formalized approach to location tracking, would have supported higher-quality measurements. Second, additional methods of documenting events of interest would help in the subsequent analysis of the data collected. Using time- and location-stamped photographs or video, for instance, would enable a more robust linkage between cause and effect observed in the collected data.

These issues, however, can easily be remedied, and the opportunities to engage the local community are a vital element of urban data science and civic technology adoption. In addition, this type of citizen science effort provides an opportunity to support workforce development and education. If data science and informatics are to be used to have a positive impact and improve quality-of-life, the earlier and more substantively the local community can be involved, the better chance of real change.

## 4. CONCLUSION

The CUSP Quantified Community research initiative continues to grow. As our work extends across several communities, we are gaining new insights into how civic technology, urban sensing, and urban informatics can be used to better understand how neighborhoods impact community and individual well-being. By combining robust scientific inquiry with community engagement, we hope to demonstrate how we can simultaneously advance the fundamental understanding of cities, while enabling positive changes in quality-of-life. Distinct from the typical Smart City rhetoric, we view the community as the ideal scale to study urban dynamics and the locus of data-driven planning and policy efforts.

Our work in Red Hook provides an early glimpse into the challenges and possibilities of urban sensing and informatics in traditionally under-served, low-income communities. Our preliminary success in Red Hook - deploying our QC sensors, collecting data through citizen science, and analyzing the data to discover actionable insights – is a function of the leadership and innovative capacity of the Red Hook Initiative. To successfully conduct this type of work in other neighborhoods, the importance of a strong local community partner to serve as a coordinator and voice of residents and businesses cannot be overstated.

The QC initiative continues to expand its sensor deployments and data collection efforts in Hudson Yards, Lower Manhattan, and Red Hook. Each neighborhood provides a very distinct context for the study of cities. Over time, the ability to compare the different neighborhoods across indicators of quality-of-life, public health, infrastructure capacity, and mobility will provide a unique platform for the long-term study of neighborhood dynamics, create new opportunities for benchmarking the quality of neighborhoods and impacts of land use and other changes, and shift the nature of city governance to encourage new innovations, rapidly evaluate their effects, and adopt ideas that are shown to be effective.

## 5. ACKNOWLEDGEMENTS

A special thanks to the CUSP Quantified Community Capstone team, including Bartosz Bonczak and Awais Malik, and CUSP M.S. students Jonathan Grundy, Eren Con, Tengfei Zheng, Clayton Hunter, Alejandro Porcel, and Maria Ortiz. The students were involved in the deployment of the QC sensors, the citizen science day, and in analyzing the data. We deeply appreciate the partnership of the Red Hook Initiative and the leadership of Tony Schloss, Director of Technology, who made the pilot project in Red Hook possible. We would also like to thank Pioneer Works in Red Hook for allowing us to install a sensor at their location.